# Tuneable local order in thermoelectric crystals

Nikolaj Roth[1], Jonas Beyer[1], Karl F. F. Fischer[1], Kaiyang Xia[2], Tiejun Zhu[2], Bo Brummerstedt Iversen[1*]

1: Department of Chemistry and iNano, Aarhus University, Aarhus 8000, Denmark
2: State Key Laboratory of Silicon Materials, School of Materials Science and Engineering, Zhejiang University, Hangzhou 310027, People's Republic of China
*Bo@chem.au.dk

**Abstract:**

Although crystalline solids are characterised by their periodic structures, some are only periodic on average and deviate on a local scale. Distinct local orderings can exist with identical periodic structures making their differences hidden to normal diffraction methods. Using x-ray scattering we investigate the thermoelectric half-Heusler, $Nb_{1-x}CoSb$, with high vacancy concentrations, where signs of local order have been attributed to differences in stoichiometry. We show that the composition is actually always very close to $x = 1/6$ irrespective of nominal sample composition, and that the synthesis method controls the local order. The vacancy distribution is shown to follow a vacancy repulsion model, and local structural relaxations around the vacancies are characterised. Control over the local structure provides a new frontier for tuning material properties.

**Introduction:**

Crystalline solids are typically understood as being periodic on the atomic scale, and are described by a repeating unit cell in three dimensions. With such an ordered arrangement of atoms, knowledge of the atomic configuration in one part of the crystal will perfectly predict the positions of all other atoms. This is in contrast to amorphous solids, where there are only local coordination rules. Knowledge of the atomic configuration at one point in an amorphous solid, does not allow prediction of atomic positions elsewhere in the material.

In general, the atomic structure of a material determines its properties. In crystals, collective behaviour of the periodic structures gives rise to delocalised modes in the electronic, magnetic and vibrational properties. For amorphous materials, the properties arise from localised behaviour. Between the two extremes we find disordered crystals, which on average can be described by an ordered periodic structure, but with deviations on a local scale. These can have more complex properties arising from both collective and localised behaviour. The average periodic structure gives rise to sharp peaks in a diffraction experiment (Bragg peaks), which can be readily analysed. The local deviations give rise to weak diffuse scattering, which is much more difficult to measure and especially to interpret on a structural basis (*1-3*).

Defective half-Heusler materials $X_{1-x}YZ$, such as $Nb_{1-x}CoSb$, have excellent thermoelectric properties and they show signs of local atomic order different from their average periodic order (*4-8*). Electron diffraction data have revealed structured diffuse scattering, which differ from sample to sample (*9*). The differences were first explained as a result of different nominal sample stoichiometry (*9*), but recently the differences in shape of the reported diffuse scattering has been theoretically explained by a simple model for effective vacancy repulsion on the disordered X substructure, without invoking different sample stoichiometry (*10*). The model suggests that the structure places vacancies as far apart as possible, essentially giving the effect of vacancy repulsion. Importantly, the model predicts that the degree of local order can be influenced by synthesis conditions, such as thermal quenching. Experimental tuning of local order would provide a new handle in materials research, which may allow for controlling properties in disordered systems.

Recently, Goodwin and coworkers used diffuse x-ray scattering data to establish vacancy distributions in Prussian Blue Analogue (PBA) materials (*3*). Different PBAs crystals showed dissimilar diffuse scattering patterns, and indeed PBA materials are known to have large variation e.g. in battery properties (*11*). However, an understanding of how local structure can be controlled and how the local structure correlates with material properties is still largely unexplored. Here, we show that the synthesis conditions change the degree of local order in thermoelectric $Nb_{1-x}CoSb$. Using diffuse synchrotron x-ray scattering data measured on single crystals, we first validate the theoretical vacancy repulsion model, and then we model the structural relaxation around the vacancies to provide direct experimental quantification of the local structure in these systems.

**Results**

The theoretical model for vacancy repulsion in defective half-Heusler materials predicts that samples quenched (Q) from high temperature will have a different degree of local order than slowly cooled (SC) samples (*10*). Using induction furnaces, two distinct types of samples have been synthesized, quenched samples from a levitation-melt procedure and slowly cooled samples using crucibles. Samples were prepared using both methods with nominal stoichiometries of $Nb_{0.81}CoSb$ and $Nb_{0.84}CoSb$, and they are named such that "Q-0.81" is a quenched sample with nominal stoichiometry $Nb_{0.81}CoSb$.

From very high-quality single-crystal x-ray scattering measurements performed at a synchrotron we obtain the average crystal structure for each single crystal from the Bragg peaks, including an accurate stoichiometry, In addition, from the same measurement we obtain the diffuse scattering, allowing for analysis of the local order. With this method we can directly correlate the diffuse scattering and local order with the periodic crystal structure and sample composition, as this is all obtained from one measurement for each single crystal. It further avoids errors in determining the sample composition using other methods, as impurity phases are known to occur in these compounds (*9*).

**Average Crystal structure**

The average periodic crystal structures in all samples were quantified by analysis of the Bragg diffraction peaks. All samples have cubic average structures in space group $F\bar{4}3m$ with the cell lengths in a narrow range between 5.894 and 5.899 Å at 300 K. The quality of the Bragg data is excellent with internal agreement ($R_{int}$) between 4.0% to 5.4% even for data resolutions with $d_{min}$ < 0.43 Å, and average redundancies of 19-36 (see Table 1). Previous studies describe the average structure as an ideal half-Heusler structure with vacancies on the Nb sites (*7*). The ideal half-Heusler structure (space group $F\bar{4}3m$) has $Nb_{1-x}$ at [0, 0, 0], Sb at [1/2, 1/2, 1/2] and Co at [1/4, 1/4, 3/4]. $Nb_{1-x}$ and Sb form a rock-salt structure, while Nb and Co form the sphalerite structure, as illustrated in Fig. 1a.

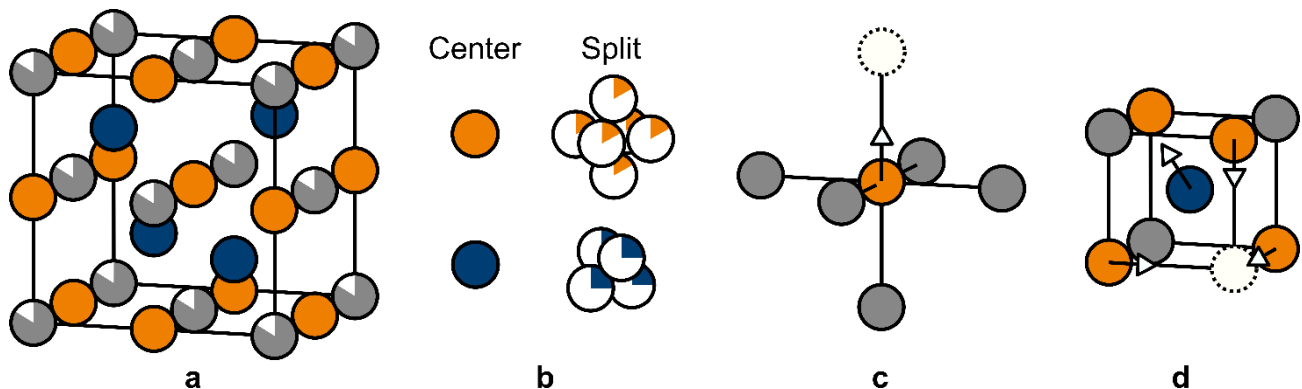

**Figure 1: Crystal Structure of $Nb_{1-x}CoSb$.** (**a**) The unit cell of the average structure. Blue is Co, grey is Nb, Orange is Sb. Partial shading indicates occupancy. (**b**) Two models have been tested, one with Sb and Co at the ideal half-Heusler positions and one with off-centered Sb and Co (exaggerated in the figure). (**c**) A $Nb_5Sb$ elementary block with the Sb displacement indicated by the arrow. (**d**) Local relaxation of Co close to a vacancy with the movements of Sb and Co indicated by arrows.

Refining a model with free Nb occupancy gives good (low) agreement factors ($R_1$: 2.29% - 3.23%, $wR_2$: 5.94% - 7.39%), which normally would be considered excellent fit, especially with the large range of data. However, clear residuals are observed around the Sb and Co sites suggesting disorder, as shown in detail in the supplementary materials. If Sb is positioned off-center at [½+Δ, ½, ½] with refinement of Δ, and Co is off-centered at [1/4-δ, 1/4-δ, 3/4-δ] with refinement of δ, the agreement factors improve significantly for all samples ($R_1$: 0.49% - 1.06%, $wR_2$: 0.78% - 1.33%), Table 1. In this model, each Sb is split into six positions (each with 1/6 occupancy) forming an octahedron with corners pointing towards the six neighbouring $Nb_{1-x}$ sites, while Co is split into a tetrahedron with corners pointing towards neighbouring Sb, see Fig. 1b. For all samples, the Sb shift refines to Δ ~ 0.025 corresponding to a movement of 0.144 Å, and Co shifts δ ~0.0127 corresponding to a movement of 0.130 Å. Interestingly, the refined Nb occupancies are in the narrow range 0.820 to 0.827 for the five samples, suggesting that all $Nb_{1-x}CoSb$ samples are very close in stoichiometry with x ~ 1/6. Thus, there is no correlation between the nominal and refined sample stoichiometry. It is remarkable the average structure of all samples is identical with x ~ 1/6 irrespective of synthesis method and nominal sample stoichiometry. In all samples, Sb is shifted off-center by around 0.144 Å and Co by around 0.130 Å. From refinements of the Bragg data all samples are identical, but "hidden" local structures are exposed through analysis of the diffuse scattering.

**Table 1: Average structure refinements for the different $Nb_{1-x}CoSb$ samples.** "Q" are quenched samples, "SC" are slowly cooled samples. The decimal number designates the nominal Nb stoichiometry. "Avg. red." is the average data redundancy, i.e. the average number of times each unique reflection was measured.

| Sample | $a$ [Å] | avg. red. | $d_{min}$ [Å] | $R_{int}$ [%] | Center model $R_1$ [%] | $wR_2$ [%] | $occ_{Nb}$ | Split model $R_1$ [%] | $wR_2$ [%] | $occ_{Nb}$ | Sb shift [Å] | Co shift [Å] |
|---|---|---|---|---|---|---|---|---|---|---|---|---|
| Q-0.81 | 5.896 | 33 | 0.33 | 4.32 | 2.27 | 6.06 | 0.827 | 0.63 | 0.83 | 0.827 | 0.142 | 0.131 |
| SC-0.81 | 5.894 | 34 | 0.43 | 4.02 | 2.28 | 6.20 | 0.835 | 0.89 | 1.16 | 0.820 | 0.148 | 0.128 |
| Q-0.84 #1 | 5.899 | 19 | 0.40 | 5.36 | 3.23 | 7.39 | 0.820 | 1.06 | 1.33 | 0.825 | 0.142 | 0.130 |
| Q-0.84 #2 | 5.896 | 36 | 0.40 | 4.28 | 2.76 | 6.00 | 0.831 | 0.49 | 0.78 | 0.827 | 0.141 | 0.130 |
| SC-0.84 | 5.895 | 29 | 0.40 | 4.16 | 2.94 | 5.94 | 0.820 | 0.98 | 1.27 | 0.822 | 0.146 | 0.131 |
| Average | 5.896 | 30 | 0.39 | 4.43 | 2.70 | 6.32 | 0.827 | 0.81 | 1.07 | 0.824 | 0.144 | 0.130 |

**Tuneable short-range order**

Although all samples have almost identical average structures, their scattering patterns are highly different. In general, the thermally quenched (Q) samples show more diffuse scattering than the slowly cooled (SC) samples, which have sharp additional peaks. Fig. 2a shows the measured scattering in the H0L and HHL planes for two representative samples, while scattering patterns from the remaining samples are in the supplementary materials. There is no correlation between the nominal sample stoichiometry and the degree of diffuse scattering, but clearly the diffuse scattering depends on the synthesis method. The inverse of the broadness of the diffuse scattering is proportional to the correlation length of local order. Thus, the quenched samples with broader diffuse scattering only have short-range order, whereas the slowly cooled samples with sharp peaks have longer range order.

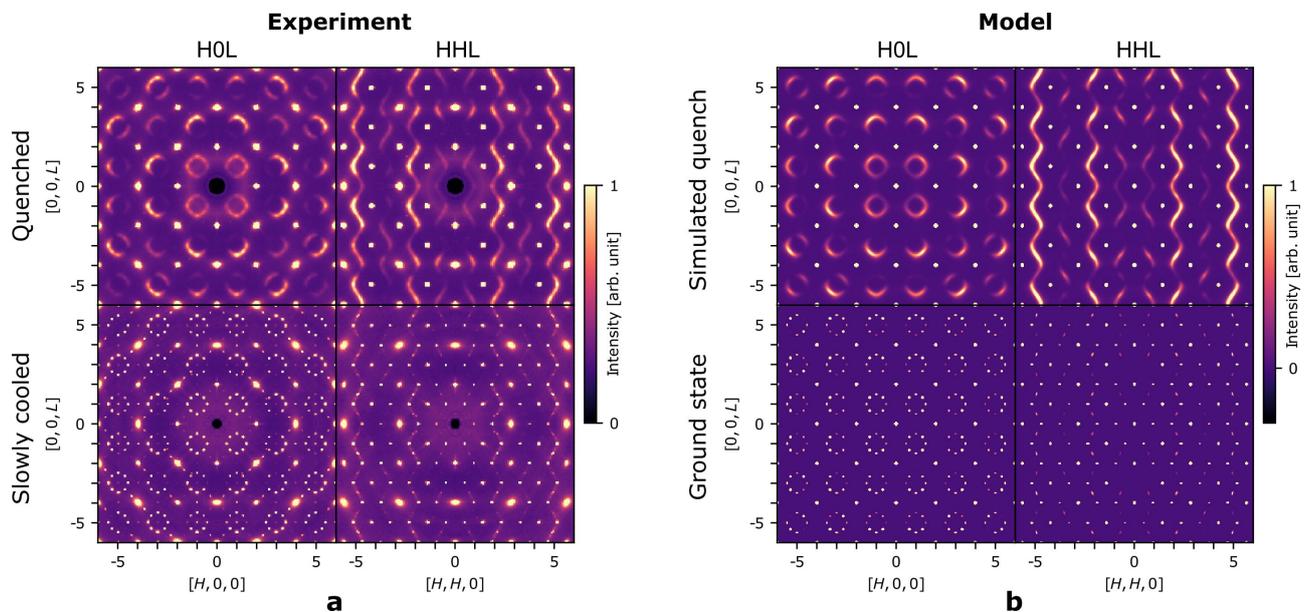

**Figure 2: Measured and calculated x-ray scattering in the H0L and HHL planes.** (**a**) Measured x-ray scattering from two representative samples of $Nb_{1-x}CoSb$ at 300 K with the top row being a quenched sample (Q), and the bottom row a slowly cooled (SC) sample. (**b**) Calculated x-ray scattering with the top row showing a simulated quench model, and the bottom showing a ground state of type "BD" (see text).

The measured x-ray scattering data, Fig. 2a, differ significantly from the electron diffraction data reported by Xia *et al*. (*9*), where the diffuse scattering consists of rings in the H0L plane with approximately constant intensity around the perimeter. In the x-ray data, the rings have clear intensity modulations. The difference is due to strong multiple scattering in the electron diffraction data, where the electron beam of strong Bragg peaks is re-scattered to give diffuse scattering averaged over Brillouin zones. This creates rings of approximately constant intensity, especially when data is measured along the zone-axis where Bragg scattering is strong as done by Xia *et al.* (*9*)

The theoretical model for the vacancy distribution (*10*) was inspired by the electron diffraction data, and it produces calculated rings without the intensity modulation observed in the present x-ray scattering data. As will be shown, the intensity modulation of the diffuse scattering is mainly the result of structural relaxation of Sb and Co around vacancies, but overall the vacancy ordering model is validated. Similar modulations due to structural relaxation around vacancies, a type of size-effect, have been reported in other compounds (*12, 13*).

To get a direct view of the local correlations in the samples, the diffuse scattering is Fourier transformed to obtain the thee-dimensional difference pair distribution function (3D-ΔPDF), which is the autocorrelation of the difference electron density:

$$3\text{D-}\Delta\text{PDF} = \langle \delta\rho \otimes \delta\rho \rangle$$

Here $\delta\rho(\boldsymbol{r},t) = \rho(\boldsymbol{r},t) - \rho_{periodic}(\boldsymbol{r})$ is the difference between the total electron density of the crystal and the periodic average electron density. The 3D-ΔPDF gives a direct view of the local deviations from the periodic average structure (*14-16*). Positive/negative features show vectors for which the real structure has more/less electron density separated by those vectors compared with the average structure. The types and signs of features can be directly related to the types of local correlations (*16*), and indeed the 3D-ΔPDF has been used to solve the local order in several disordered crystals (*2, 17, 18*).

Fig. 3a and 3b show the 3D-ΔPDF in the 010 plane. The short-range features are almost identical in the two representative samples, but the features decay quickly in the quenched sample while they have longer range for the slowly cooled sample. This is illustrated in more detail in the supplementary materials. In general, the 010 plane is dominated by features which are positive on one side and negative on the other, a direct indication of strong local relaxation around vacancies (*16*).

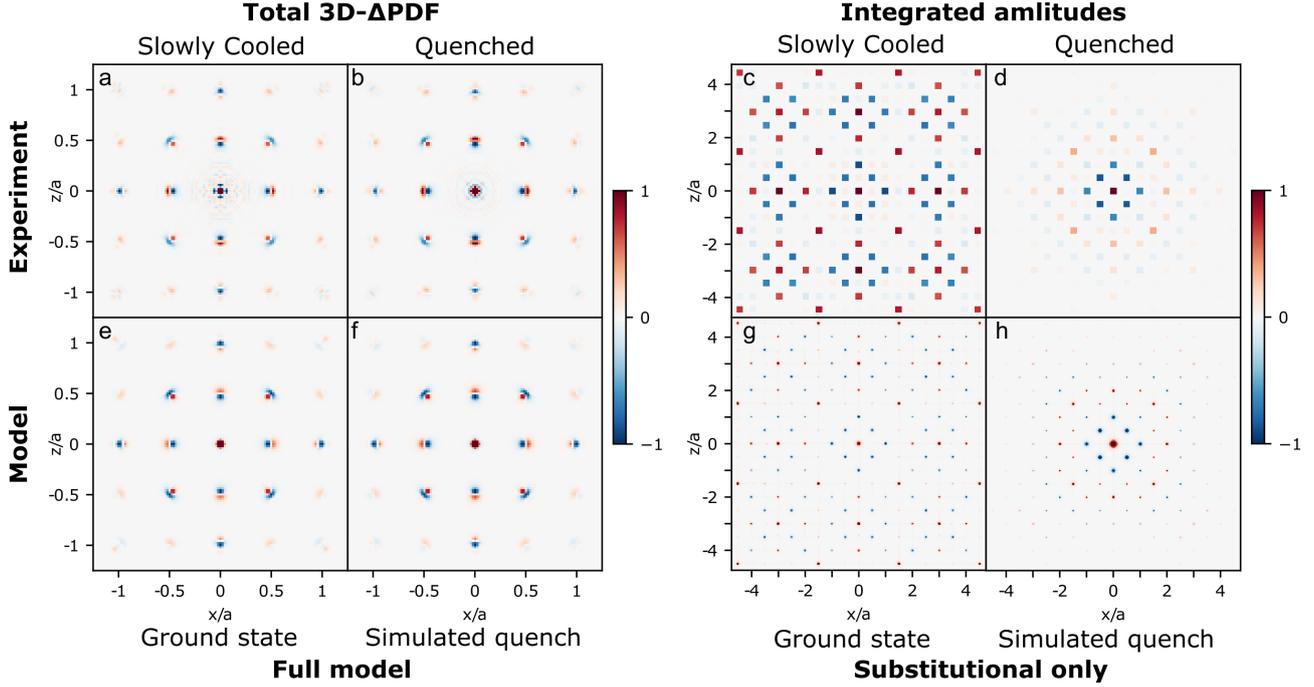

**Figure 3: Measured and calculated 3D-ΔPDF**. (**a**) and (**b**) show the measured 3D-ΔPDF for the "SC-0.81" and "Q-0.84 #2" samples in the 010 plane (same samples as in Fig. 2). (**c**) and (**d**) show the integrated peak amplitudes. (**e**) and (**f**) show the calculated 3D-ΔPDF using the vacancy repulsion model *including* Sb and Co relaxation. Both plots have a vacancy concentration of x = 1/6 but (e) represent a BD-type ground state and (f) is for a simulated quench with $J_2/J_1$=0.5. Details of the ground state structure is given in the supplementary materials. (**g**) and (**h**) show the calculated 3D-ΔPDF using the vacancy repulsion model without Sb and Co displacements.

**Vacancy ordering**

To remove the effect of atomic movements and isolate the effect of local vacancy ordering, the features in the 3D-ΔPDF can be integrated. As shown by Roth *et al.* (*17*), the integral amplitudes of features in the 3D-ΔPDF relate only to the substitutional disorder. The integrated peak amplitude of a peak at position $\boldsymbol{r'}$ is proportional to

$$\text{peak amplitude at } \boldsymbol{r'} = \sum_{(i,j)|\boldsymbol{r}_{ij}=\boldsymbol{r'}} \delta Z_i \cdot \delta Z_j$$

, where the summation runs over all pairs of atoms $(i,j)$ separated by vector $\boldsymbol{r}_{ij} = \boldsymbol{r'}$, and $\delta Z_i$ is the difference in number of electrons of atom $i$ in the real structure compared to the average periodic structure.

It is possible to use this method on the current samples as the features in the 3D-ΔPDF are well-separated. Details of the peak integration is shown in the supplementary materials.

The integral peak amplitudes are shown in Figure 3c and 3d. A positive value is found at the origin since atoms are separated by the zero vector to themselves. Surrounding the origin are strong negative integral amplitudes corresponding to nearest and next-nearest vectors in the Nb substructure (coordinates (½,½,0) and (1,0,0) ). This shows that there is a lower-than-average probability of finding two atoms separated by these vectors. This means that the structure tends to avoid nearest and next-nearest vacancy pairs. Then at slightly longer distances there are several positive amplitudes showing the preferred distances between vacancies. Again, the slowly cooled samples have strong correlations to much longer distances than the quenched samples.

The integral amplitudes can then be directly compared to the calculated 3D-ΔPDF obtained from a simulation of the vacancy distribution using the vacancy repulsion model (10). The model gives a positive energy to each nearest and next-nearest vacancy pair with the total energy of the system given by

$$E = J_1 N_1 + J_2 N_2$$

Here, $N_1$ is the number of nearest-neighbour vacancy pairs and $N_2$ is the number of next-nearest-neighbour vacancy pairs on the Nb substructure. $J_1$ and $J_2$ are the energy penalties of nearest and next-nearest neighbour vacancy pairs. The vacancy concentration x = 1/6 is the highest vacancy concentration where it is possible to avoid all nearest and next-nearest neighbour vacancy pairs, as was shown in (10). The ground states for the model are therefore all configurations with no such pairs (E = 0). There is a large number of configurations satisfying those rules, and different possible types were previously identified (10). The integrated amplitudes for the slowly cooled samples agrees with the calculated 3D-ΔPDF for a vacancy model in the ground state, as shown in Fig. 3g. The amplitudes are in the best agreement with the ground state type "BD" (details in the supplementary materials).

For the non-ground-state configurations, a Monte-Carlo simulation can be carried out using the Metropolis algorithm (19). Steps with a positive energy change are accepted with probability $P = e^{-\frac{\Delta E}{T}}$, where T is the simulated temperature. To simulate the quenched samples, a configuration with x = 1/6 is started from random vacancy positions, and the simulation is run for T = 0. The relative energy penalty for next-nearest vacancy pairs was set to half the energy penalty of nearest pairs, $J_2/J_1 = \frac{1}{2}$. The resulting simulated 3D-ΔPDF is in good agreement with the integrated amplitudes from the quenched samples, as shown in Fig. 3h.

For both the quenched and slowly cooled samples, the vacancy correlations from the simulation correspond well with those obtained by integration of amplitudes in the measured 3D-ΔPDF. This shows that the model for vacancy correlations (*10*) agrees with the experimental data, even though the model does not reproduce the intensity modulations of the diffuse scattering rings.

**Structural relaxation around vacancies**

The measured 3D-ΔPDF (Fig. 3a and 3b) show strong features from local structural relaxation around vacancies. At (x,y,z) = (½,0,0), the 3D-ΔPDF is negative toward the center and positive away from the center. This is the vector between Nb/vacancies and Sb. This means that when a Nb is present, the Sb will be slightly further away, and when a vacancy is present, the Sb will move towards the vacancy position.

In the average structure each Sb is surrounded by 6 close Nb/vacancy positions, forming an octahedron. Avoidance of nearest and next-nearest vacancy requires each such octahedron to have at most one vacancy out of 6 corners. For x=1/6 there will be one and only one close vacancy per Sb in the ground state. This means that each Sb simply moves towards the one vacancy neighbour it has by approximately 0.144 Å. Consequently, the (Nb,Sb) substructure can be seen as being built from $Nb_5Sb$ square pyramidal units, where the Sb is displaced slightly towards the empty Nb site, as illustrated in Fig. 1c.

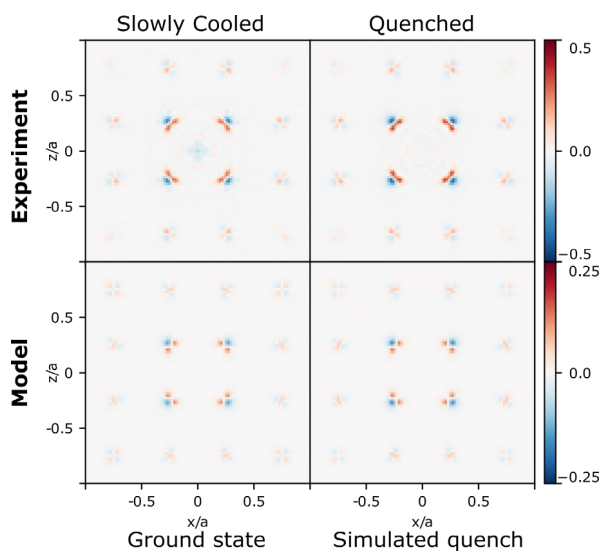

**Figure 4: Measured and calculated 3D-ΔPDF in the y = 0.27a plane.** This plane shows vectors which are all related to correlations between Co and other atoms. The top row shows the measured 3D-ΔPDF for the two representative samples, while the bottom row shows the corresponding 3D-ΔPDF calculated from the vacancy repulsion model with structural relaxations. Note that the experiment and model 3D-ΔPDF are shown on different scales. The measured 3D-ΔPDF has stronger amplitudes, indicating the model to still have something missing.

The relaxation of Co close to vacancies can also be identified from the 3D-ΔPDF. The top row in Fig. 4 shows the (x, y = 0.27a, z) plane of the 3D-ΔPDF for the two representative samples. This plane shows strong features which are all related to correlations between Co and the other atoms. Each Co sits in a cube formed by Sb and $Nb_{1-x}$. The shortest Co-Nb and Co-Sb vectors, which are identical in the average structure, are located around (0.25, 0.25, 0.25). As seen in the top of Fig. 4, the feature is mostly positive towards the origin and negative away from the origin. This means Co moves away from a vacant Nb site, consistent with the tetrahedron of Co sites in the average structure. This is illustrated in Fig. 1d where the local environment of a Co close to a vacancy is shown. Since the local relaxation of both Sb and Co has been identified from both the average structure and the 3D-ΔPDF, the local structure model can be improved and compared with the measurements.

To improve the theoretical model, the vacancy distributions obtained from the Monte-Carlo simulations are used, but with each Sb moved by 0.144 Å toward its neighbouring vacancy, and with Co moved by 0.130 Å away from neighbouring vacancies. The calculated 3D-ΔPDFs for this model are shown in Fig. 3e and 3f for the (010) plane. These are in good agreement with the experimentally obtained 3D-ΔPDF maps, further validating the local structural relaxation around vacancies. The bottom row of Fig. 4 shows the calculated 3D-ΔPDF in the y = 0.27a plane and a good agreement with experiment is observed. However, the amplitudes of the features are significantly weaker for the model than for the experiment suggesting still undescribed features in the model. This could be small relaxations of Nb in its local environment, or possibly more complex movements of Co/Sb than investigated here.

The calculated scattering for the models is shown in Fig. 2b. The intensity modulations of the diffuse scattering is now to a large extend reproduced by the model. Both the short-range order of the distribution of Nb/vacancies and the displacements of Sb and Co are needed to explain the measured data. A previous study claimed that the Nb/vancancy ordering was insignificant to the diffuse scattering signal (*20*) based on erroneous analysis, as explained in the supplementary materials. There are still small differences in the measured and calculated scattering. For example, there are weak additional peaks for the slowly cooled samples not currently reproduced by the model. This is discussed in the supplementary materials.

**Discussion**

The quenched samples show more diffuse-like scattering than the slowly cooled samples, which have sharp additional peaks. This shows that the degree of local order can be tuned through the synthesis conditions (*10*). Surprisingly, the refined chemical compositions are virtually identical in all samples, x≈1/6, even though the nominal stoichiometry used in the syntheses differ substantially. Accurate determination of the composition was made possible by the high-quality single-crystal x-ray scattering data, avoiding errors from other impurity phases known to occur in these systems (*9*). This suggests that the stable composition is $Nb_{5/6}CoSb$, and that there is very little room for variation in composition. Previous studies have argued that the ideal composition is $Nb_{4/5}CoSb$ based on simple valence electron counting (Zintl) rules (*7*). However, x = 1/6 is the highest vacancy concentration for which both nearest and next-nearest vacancy pairs can be avoided completely (*10*). This means that the energetic penalty for disturbing the local chemical bonding in the locally ordered vacancy structure is higher than the presumed gain in electronic energy expected from simplistic electron counting. It is noteworthy that the average structure is identical for all five samples, and also the local structural relaxation is the same with Sb is being off-centered by 0.144 Å and Co off-centered by 0.130 Å. Each Sb has one vacant neighbour and moves toward it, while Co moves away from neighbouring vacancies. Both the

Having established that all samples have x ~ 1/6, but differ in their local structure, a key question is how the local structure affect the thermoelectric properties. It is difficult to measure thermoelectric properties on samples with controlled local structure, since fabrication of a pellet from polycrystalline material requires densification under heat and pressure. However, Xia *et al.* reported thermoelectric properties of six samples with x varying between 0.15 and 0.20, and all synthesized by the quench method. The samples had identical lattice parameters (from PXRD) strongly corroborating that x = 1/6 for all samples as suggested here. The differences in carrier concentrations are likely affected by the necessary impurities present when nominal compositions deviate from 1/6. The measured transport data strongly reflect the presence of vacancies in their temperature behaviour, but presumably it is the lattice thermal conductivity, $\kappa_L$, which most directly is affected by the local structure. At room temperature, the samples differ by about 15% in $\kappa_L$ and this could indeed be a direct effect of different vacancy distributions. It could therefore be advantageous attempt to use these differences in properties. Electron-diffraction data showed the diffuse features to be stable even at around 1000K, suggesting the vacancy distributions to be stable to quite high temperatures (*9*), making it possible to make devices with a specific vacancy order.

In summary, we have shown that defective half-Heusler materials $Nb_{1-x}CoSb$ have a strong tendency to vacancy ordering following a simple repulsion model. The optimal ordering of vacancies essentially fixes the stoichiometry of the samples irrespective of the nominal starting composition or synthesis method. Using analysis of both Bragg diffraction and diffuse scattering data with the 3D-ΔPDF method, it is possible to quantify the structural relaxation around a vacancy site. Different local structure states are reached depending on the thermal treatment of the sample, and they appear to have appreciable effect on the thermoelectric properties. Advanced x-ray scattering techniques can unravel hidden local structures and for $Nb_{1-x}CoSb$ these local structures can be controlled by the synthesis conditions. If the local structure of crystalline materials can be more generally related to the properties, then a new frontier in materials research will be available.

**Acknowledgments**:

The authors gratefully acknowledge Martin von Zimmermann for sharing his script for converting raw data to reciprocal space. Kunihisa Sugimoto, Kristoffer A. U. Holm, Kasper Tolborg, Thomas Bjørn Egede Grønbech and Lennard Krause are thanked for helping carry out the synchrotron experiments. Affiliation with the Aarhus University Center for Integrated Materials Research (iMAT) is gratefully acknowledged. The synchrotron experiments were performed at SPring-8 BL02B1 with the approval of the Japan Synchrotron Radiation Research Institute (JASRI) as a Partner User.

**Funding:**

Villum Foundation

Danish Agency for Science, Technology and Innovation, DanScatt

**Supplementary Materials for**

# Tuneable local order in thermoelectric crystals

Nikolaj Roth[1], Jonas Beyer[1], Karl F. F. Fischer[1], Kaiyang Xia[2], Tiejun Zhu[2], Bo Brummerstedt Iversen[1*]

1: Department of Chemistry and iNano, Aarhus University, Aarhus 8000, Denmark

2: State Key Laboratory of Silicon Materials, School of Materials Science and Engineering, Zhejiang University, Hangzhou 310027, People's Republic of China

**Synthesis**

Two types of synthesis methods were used to produce samples with nominal stoichiometries $Nb_{0.81}CoSb$ and $Nb_{0.84}CoSb$. One group of samples were thermally quenched from the melt (levitation-melt technique), while the other group of samples were slowly cooled using an induction furnace.

The quench samples are of the same type used in the published electron diffraction study (*9*).

Slowly cooled (SC) samples of compositions $Nb_{0.81}CoSb$ and $Nb_{0.84}CoSb$ were prepared from stoichiometric mixtures of the pure elements with a 1% molar excess of Sb to compensate for the evaporation of it during synthesis. Pieces of Sb and Co, and wire clippings of Nb were loaded into an alumina crucible, which was then slotted into a graphite susceptor inside an induction coil. The furnace chamber was evacuated thrice to a level between 0.01 and 0.1 mbar with a back filling of He in between evacuations, before being pressurized to 10 bar with He. The output power of the furnace was turned up in large steps over a 5 min period to the point where a homogeneous melt is achieved. Based on the light emitted from the susceptor this occurs at a temperature between 1400 and 1500 °C. After a 1 min soak time at the maximum temperature, the output power was gradually decreased over the following 5 min until the whole melt had solidified, at which point the temperature of the susceptor was around 1100 °C. After the sample had solidified, the furnace was turned down and finally off over a period of 5 min, after which the sample was left in the furnace chamber for 10 min to cool to room temperature. Comparing the mass of the used materials and the obtained samples shows that only between 0.015 and 0.02 g are lost during synthesis of a 10 g sample, which corresponds to only between ⅓ and ½ of the added excess Sb.

**X-ray scattering measurements**

Data were collected on five samples: quenched Nb$_{0.81}$CoSb (Q-0.81), slowly cooled Nb$_{0.81}$CoSb (SC-0.81), two quenched Nb$_{0.84}$CoSb samples (Q-0.84#1 and Q-0.84#2) and one slowly cooled Nb$_{0.84}$CoSb (SC-0.84). Single crystals of sizes 50-80 μm were glued to the end of thin glass pins using epoxy and mounted on a goniometer. Data was measured at the BL02B1 beamline at the SPring-8 synchrotron using a photon energy of 50.00 keV on a Huber four-circle (quarter chi) goniometer equipped with a Pilatus3 X 1M CdTe (P3) detector. For the quench samples a detector distance of 130 mm was used, while it was 260 mm for the slowly cooled samples to properly separate the sharp additional peaks.

Two measurements were made for each sample. First, a dataset for the strongest reflections were collected. For this, a 600μm Ni film was used to attenuate the beam to 31% to avoid problems of too high flux on the detector. Here 4 runs were measured, each a 180° ω rotation with 900 frames, for χ=0° and χ =45° with the detector at 2θ=0° and 2θ =-25°. An exposure time of 0.8 seconds per frame was used. Then the weak scattering was measured without any beam attenuator. Here 6 runs were measured, each a 180° ω rotation with 900 frames, χ=0° and χ =45° with the detector at 2θ=0°, 2θ=-12.5° and 2θ =-25°. An exposure time of 1.6 seconds per frame was used. Finally the background and air-scattering was measured using the same set of exposure times, detector positions and beam attenuation as for the crystal measurement. For each combination of these, 200 frames of air scattering were measured.

To obtain the average structure, the images were converted to the Bruker .sfrm format (*21*) and the Bragg peaks were integrated using SAINT (*22*). The integrated data were processed and corrected using SADABS (*23*) and merged using SORTAV (*24*). Initial structure solution and refinement was carried out with SHELXS and SHELXL, using the Olex2 GUI (*25, 26*), with subsequent structure refinement using Jana2006 (*27*). The space-group is $F\bar{4}3m$. Anomalous scattering factors for 50keV were used, as implemented in Jana2006.

For the diffuse scattering analysis the data was converted to reciprocal space using a custom Matlab script. During this process the data was corrected for polarization, the background scattering from air was subtracted, and a solid angle correction was applied. The resulting data was symmetrized using the $m\bar{3}m$ point symmetry of the Laue group. The resulting scattering data was reconstructed on a 901 x 901 x 901 point grid with each axis spanning $\pm 27$ Å$^{-1}$ for the quenched samples and $\pm 21$ Å$^{-1}$ for the slowly cooled furnace samples.

For the production of a 3D-ΔPDF, the Bragg peaks of the average structure were punched and filled. Because the Bragg peaks and diffuse scattering do not overlap, the Bragg peaks were removed using a spherical punch based on the positions of allowed reflections of $F\bar{4}3m$. The punched holes were then filled using a 3D spline interpolation. The resulting data containing only the diffuse scattering/additional peaks was then Fourier transformed to give the 3D-ΔPDF.

**Simulated data**

Monte-Carlo simulations of the structure were performed using the Metropolis algorithm (*19*) using a custom python script. The model scattering was calculated using the Scatty software (*28*), and model 3D-ΔPDF obtained by Fourier transforming the model scattering using a custom python script.

**Residual density in the average structure refinements**

Fig. S1 shows the residual electron density, calculated as an $F_{obs}$-$F_{calc}$ Fourier map for the two models of the average structure. Both the [x,0.5,z] and [x,-x,z] planes are shown. The upper row illustrates the positons of atoms in these planes. There are clear residuals around Sb and Co for the center model, showing the models to be incomplete. Because of the incomplete description, there are also spurious features at positions with no atoms in the structure, e.g. in the [x,-x,z] plane around [0.25,-0.25,0.75].

After refinement with the split site model, the residuals disappear, including the spurious features, leaving a flat and featureless difference map.

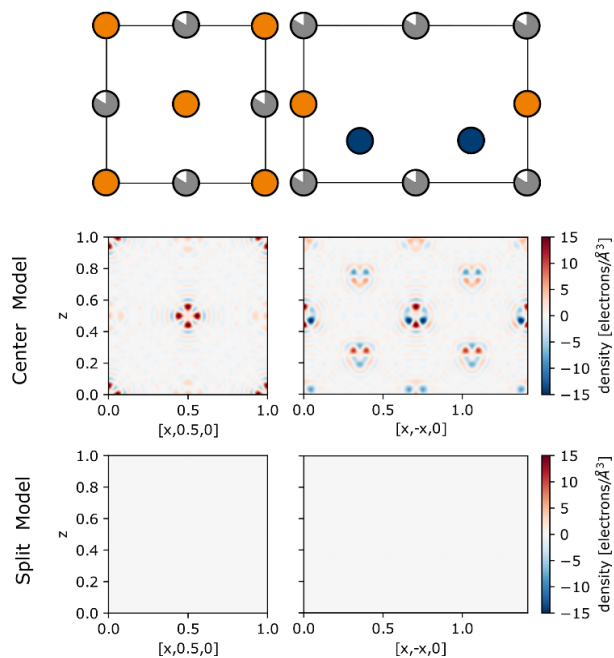

**Figure S1.** Residual electron density after refinement of the two models for the average structure. The left column shows the [x,0.5,z] plane, while the right column shows the [x,-x,z] plane. The upper row shows the ideal half-Heusler positions of atoms in these two planes. The middle row shows the residual electron density for the center model, while the bottom row is for the split model.

**Measured X-ray scattering**

The measured X-ray scattering in the H0L and HHL planes are shown below for all samples.

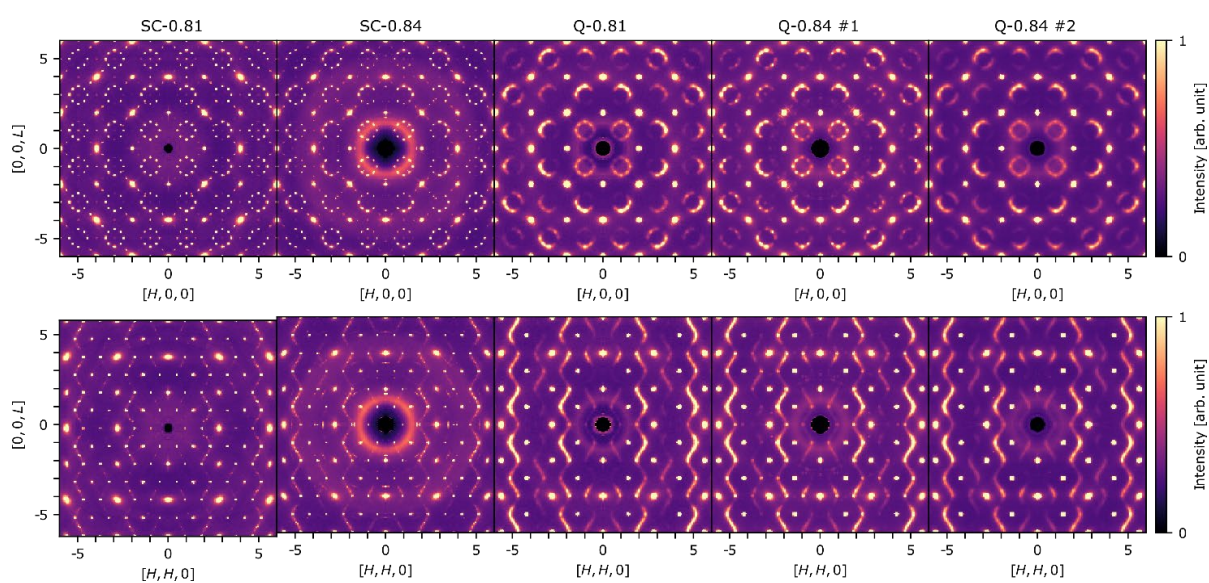

**Figure S2:** Measured X-ray scattering in the H0L and HHL plans for the different samples.

For the levitation-melt synthesis with nominal stoichiometry Nb$_{0.84}$CoSb, data were measured on two crystals ("Q-0.84 # 1" and "Q-0.84 \# 2"). While crystal "Q-0.84 \# 2" shows quite broad diffuse scattering, crystal "Q-0.84 # 1" has sharper additional peaks. This suggests that the levitation-melt samples have inhomogeneities in the degree of short-range order, which might occur as different parts of the sample are cooled at different rates during the quenching process.

**Integrated peak amplitudes**

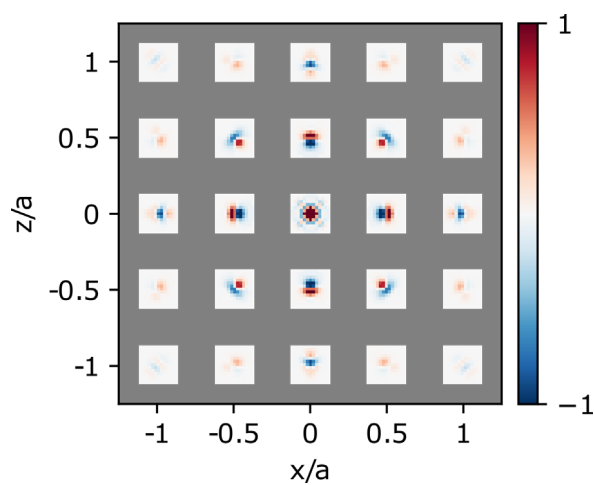

**Figure S3:** Two-dimensional view of the 3D integration boxes around features.

Peak integration is done by summing all voxels of the 3D-ΔPDF in a three-dimensional box around each feature. The used boxes are illustrated in Figure S3. Here the "Q-0.84 #1" sample is used to illustrate the method.

**Longer data ranges**

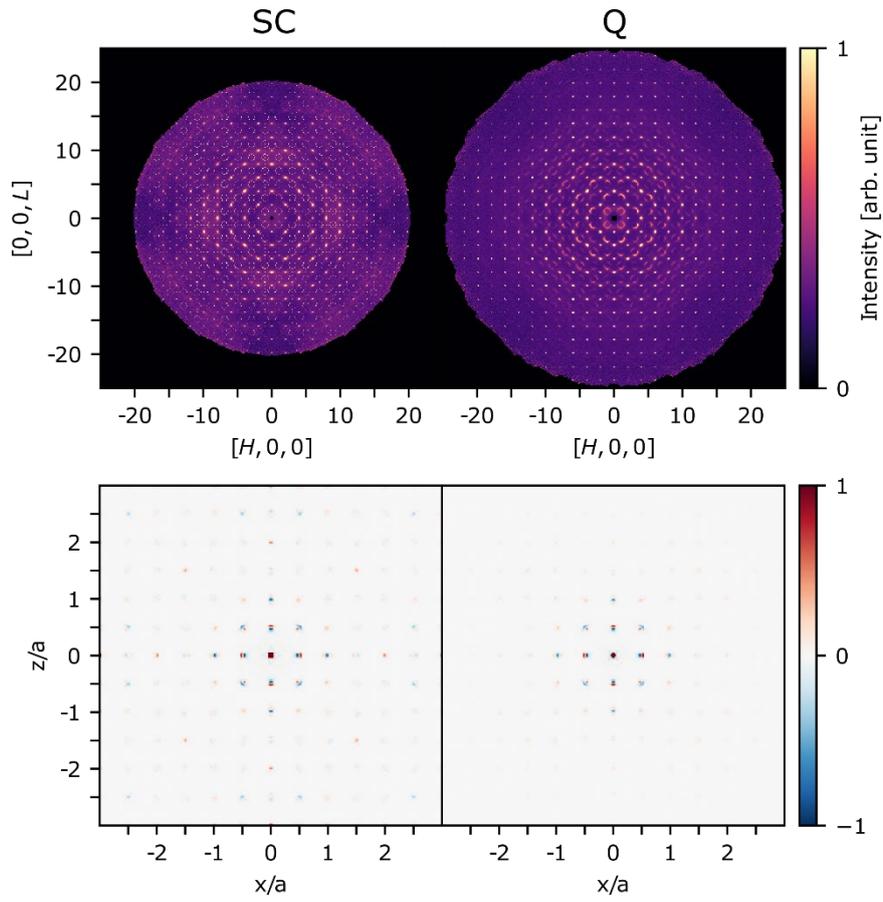

**Figure S4.** The measured scattering in the H0L plane and the 3D-ΔPDF in the 010 plane for the two representative samples shown with a larger range.

**Different ground state candidates**

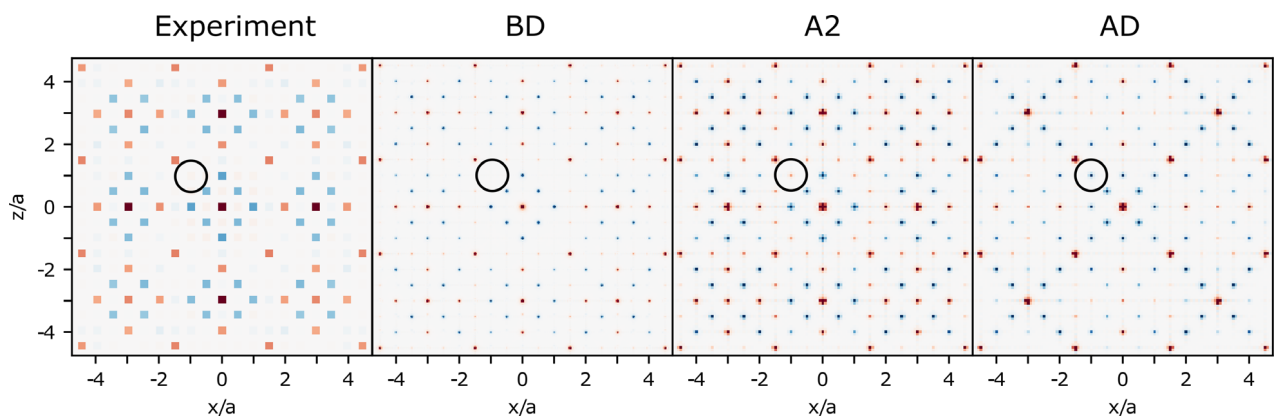

**Figure S5.** Integrated 3D-ΔPDF for experimental data compared with calculated 3D-ΔPDF for different ground state candidates.

In the vacancy repulsion model, vacancies on the Nb substructure avoid nearest and next-nearest pairs. For $x = 1/6$ there are several types of vacancy configurations that completely avoid nearest and next-nearest pairs (*10*). Of these, three have scattering patterns consistent with the measurement, as was shown previously (*10*). These are the A2, AD, and BD type ground states of the model. AD and BD are non-periodic and disordered.

The left panel of Fig. S5 shows the integrated amplitudes from the measured data on the SC-0.81 sample, which had the sharpest additional peaks in the scattering pattern. The three panels to the right show the calculated 3D-ΔPDF maps in the 010 plane for the three candidate ground states of the vacancy repulsion model without Sb and Co relaxations.

Both the A2 and BD models are in general good agreement with the experiment, while the AD model has several discrepancies in the sign of amplitudes. The black circles marks a feature which is different for the three models. For the BD model it is zero, for the A2 model it is positive and for the AD model it is negative. In the experiment this integrated amplitude is approximately zero, but weakly positive, suggesting the "BD" model to be the most accurate. However, there are still some differences between the measurement and BD model, suggesting either the BD model is not the complete description, or that the peak integration method is not accurate enough.

**Calculated scattering**

Fig. S6 shows the scattering in the H0L and HHL planes for one of the induction furnace samples together with several ground state models with vacancy repulsion and relaxation of Co and Sb. The top row shows the measured scattering in the SC-0.81 sample. The second row shows the calculated scattering for the BD ground state. Many of the features observed in the measurement are also reproduced by the model. This includes the 8 point rings of sharp scattering and some of the intensity modulations. The intensity modulation of the rings marked by green, red and blue circles are reproduced quite well. However, there are also clear differences when comparing these. The measured data shows peaks inside the 8-point rings, which is not in the current model scattering. See for example the ring marked by the white circle. As was shown in the previous paper (*10*), the peaks inside the ring can be explained by the vacancy repulsion model for $x > 1/6$ and $J_2/J_1 < 0.5$. The third row in Fig. S6 shows the calculated scattering for a simulation with $x = \frac{1}{5}$, and $J_2/J_1 = 0.45$. This produces the additional peaks inside the rings. However, they are much weaker than in the experiment and not very visible in the figure. The fourth row shows the same simulated data with a 10% colour range compared to the third row to visualize the additional peaks better. The weakness of the peaks together with the following discussion suggest that the additional peaks in the experiment come from a different origin. The SC samples refine to stoichiometries of $Nb_{0.822}CoSb$ and $Nb_{0.820}CoSb$, suggesting the deviation from $x = 1/6$ to be quite small, if significant at all. With such small deviations the model does not produce clear peaks at the ring center. Furthermore, the peaks inside the 8-point rings in the measured data seem to become stronger with increasing q, while they decreases with q in the simulated data (see the blue, yellow and white circles). This could suggest that there is another reason for the peaks inside the rings. There are also further weak peaks in the measured scattering from the SC sample, not currently explained by the model, showing that there is still more to learn about these samples.

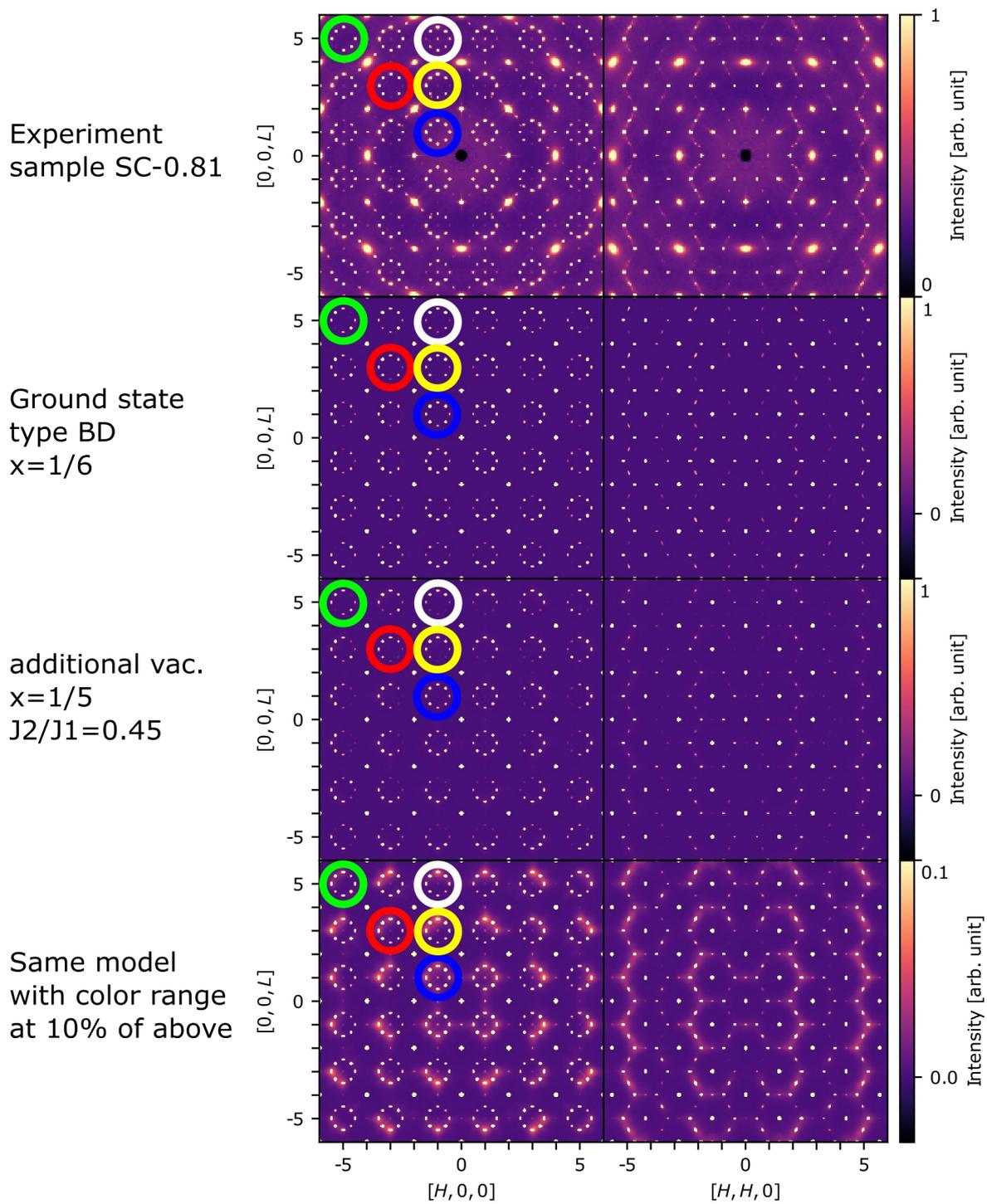

**Figure S6.** Experimental data and calculated scattering for several simulated ground-state models.

**Comments on the paper by Nan et al. (*20*)**

A study by Nan et al. looks at high-angle annular dark-field imaging (HAADF), a type of high-resolution scanning transmission electron microscopy (STEM) measured on a sample of nominal composition $Nb_{0.8}CoSb$ (*20*). The HAADF is measured along the [110] zone axis where there is no blocking of elements of different kinds. Based on this it identifies that three different amplitudes correspond to the three different elements. It then tries to show that the short-range order is mainly displacive by making Fourier transforms of the measured and ideal amplitudes together with the measured and ideal positions of atoms.

The problem is that HAADF cannot see the vacancies. The transmission-type measurement gives a projection down along the 110 zone axis. This means that when a vacancy is present on a Nb position, the measurement will still see the Nb from other layers, as not all Nb down along a column will be missing. This is clear when looking at figure 2b in their paper. Clearly a dot is seen for every Nb position, as it is a projection along 110, giving an average over many layers. Because they cannot see the vacancies but only some degree of the displacements, they reach the conclusion: "*We found that the short-range order is predominantly displacive.*" However, this is meaningless as their measurement does not allow analysis of the compositional disorder.

If the short-range order in the ED pattern was predominantly displacive, it would also not be compatible with the measured ED pattern (fig. 1a in their paper). Displacive short-range order gives diffuse scattering which is very weak close to the beam center and increases away from the beam center, whereas compositional short-range order gives diffuse scattering which is more constant in intensity (for x-ray scattering it will decrease slowly away from the beamcenter because of the atomic scattering factor). This is illustrated in figure S7.

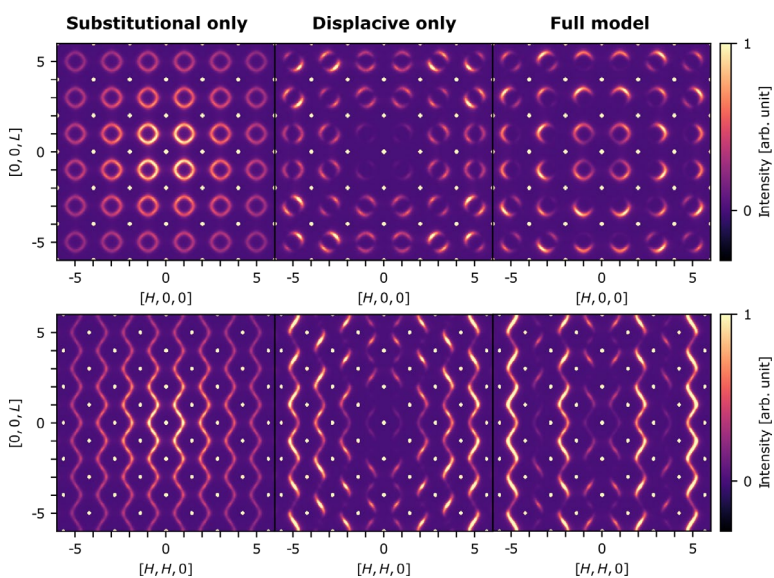

Figure S7. Calculated scattering for substitutional only (left), displacive only (center) and combined models (right).

The left column of fig. S7 shows the calculated diffuse scattering for the Nb/vacancy order, without any movements of Sb and Co. This gives diffuse scattering which decreases slowly in intensity due to the x-ray scattering factor. The central column shows diffuse scattering with only the Sb and Co displacements without any contribution from Nb. This gives diffuse scattering which is very weak close to the center and increases away from the center. The right column shows the full model with both the Nb/vacancy order and Sb and Co displacements, giving the best agreement with experiment.

In addition, the paper (*20*) avoids mentioning that the electron-diffraction data had already been explained to a higher accuracy using just substitutional short-range order (*9*), even though the authors were aware of this, as evident from them citing the paper in another context, and re-using several references that were first linked to the $Nb_{1-x}CoSb$ system in that paper.